\documentclass[pdflatex,sn-mathphys-num]{sn-jnl}

\usepackage{graphicx}%
\usepackage{multirow}%
\usepackage{amsmath,amssymb,amsfonts}%
\usepackage{amsthm}%
\usepackage{mathrsfs}%
\usepackage[title]{appendix}%
\usepackage{xcolor}%
\usepackage{textcomp}%
\usepackage{manyfoot}%
\usepackage{booktabs}%
\usepackage{algorithm}%
\usepackage{algorithmicx}%
\usepackage{algpseudocode}%
\usepackage{listings}%

\newcommand{\bra}[1]{\langle #1|}
\newcommand{\ket}[1]{|#1\rangle}

\newcommand{\braket}[2]{\langle #1|#2\rangle}
\newcommand{\ketbra}[2]{| #1 \rangle \langle #2 |}
\newcommand{\Tr}{\mathrm{Tr}}

\begin{document}

\title[Internal Interactions Cause Exponential Decay]{Bound State Internal Interactions as a Mechanism for Exponential Decay}

\author*{\fnm{Peter W.} \sur{Bryant}}\email{pwbryant@jsu.edu}

\affil{\orgname{Jacksonville State University}, \orgaddress{\street{700 Pelham Rd.~North}, \city{Jacksonville}, \postcode{36265}, \state{AL}, \country{USA}}}

\abstract{
We hypothesize that the binding interactions among the components of 
bound systems and the background fields, sometimes known as virtual particle exchange,
affect the state of the systems as do typical scattering interactions.
Then with the assumption that the interior environment of unstable particles is disordered,
we derive in the limit of
continuous binding both an exactly exponential non-decay probability and Fermi's
Golden Rule for the decay rates.
The result suggests resolutions to several long-standing theoretical challenges associated with
exponential decay in quantum mechanics,
without appealing directly to non-Hermitian, approximate Hamiltonians or complex energies.
It also contributes to a conceptual understanding of the continuum 
between controlled interactions that induce deviations from exponential decay,
such as those in the Quantum Zeno Effect,
and the uncontrolled internal dynamics of excited atoms and nuclei, which exhibit no such deviations.
Finally, we examine how the binding interactions responsible for
the general exponential character of decay for bound systems
differ from the couplings with decay products that control decay rates,
providing insight into challenges in quantum computing and information processing.
}


\maketitle

\section{Introduction}\label{sec1}
Excited quantum systems and their decay processes impact fields ranging from healthcare and
energy production to quantum computing.
Though recent advances have renewed focus on the foundations of an exponential decay
law in quantum mechanics, there remain unsettled theoretical and observational issues.
For example, there exist no exponentially decaying vectors in Hilbert spaces,
so researchers have considered non-Hermitian Hamiltonian operators, commonly
used to approximate the dynamics of open systems interacting with
environments~\cite{ashidaNonHermitianPhysics2020}.
Alternatively, Gamow states corresponding to complex energy poles 
provide exponential time evolution but also 
lead to unbounded probability distributions~\cite{bohmQuantalTimeAsymmetry2008}.
Furthermore, even if a state vector with exponential time evolution could be found,
the non-decay probability of the system it represents
cannot be exponential at large times because the 
associated Lorentzian energy wave function would contradict constraints on the
stability of matter~\cite{l.a.khalfinContributionDecayTheory1958}.

Once regarded as uniquely exponential, 
observations now suggest a continuum of decay behavior.
Deviations from exponential decay were predicted as the
Quantum Zeno Effect (QZE)~\cite{misraZenosParadoxQuantum1977} and subsequently 
demonstrated~\cite{wilkinsonExperimentalEvidenceNonexponential1997,fischerObservationQuantumZeno2001,itanoQuantumZenoEffect1990,streedContinuousPulsedQuantum2006}
for controlled, engineered systems frequently or continuously monitored to see if they had survived as they
were initially prepared.
Luminescence decays in dissolved organic molecules have revealed a deviation 
at large times, following~\cite{l.a.khalfinContributionDecayTheory1958},
where observability of the deviation requires interaction with a solvent,
and the intensity of deviations depends on internal degrees of freedom and
molecule length~\cite{rotheViolationExponentialDecayLaw2006}.
For uncontrolled bound systems, such as excited nuclei and atoms, however, deviations from
exponential decay have been sought but never
observed~\cite{ramirezjimenezQuantumDecayLaw2019,normanTestsExponentialDecay1988a}.

This continuum hints that exponential decay may not be fundamental,
but instead emerges from interactions between systems.
We therefore propose a model for bound systems that treats the internal interactions
between system components, mediated by virtual particle exchange, no differently from any scattering
or measurement interaction.
It has indeed been argued that quantum fields play a fundamental role in
measurement interactions~\cite{jaegerMeasurementFundamentalProcesses2015}.
And scattering theory has been applied to bound systems in the past,
but efforts have been hindered by the non-perturbative nature of their 
dynamics~\cite{salpeterRelativisticEquationBoundstate1951,hammerNuclearEffectiveField2020,yerokhinLambShift12015}.
Our approach is similar in spirit,
but it simplifies the treatment by focusing only on interaction kinematics.

In the continuous limit of a sequence of binding
interactions between the components of metastable, bound systems,
we derive an exactly exponential non-decay probability.
Calculating decay rates results in a derivation of
Fermi's Golden Rule (FGR) that is straightforward compared to typical textbook derivations.
We find also that the Lorentzian line-shape is emergent and does not conflict
with stability constraints.
Furthermore,
the model emphasizes the difference between internal interaction potentials and
the potentials coupling to decay products,
which is significant for the development of quantum technology.

We introduce conceptual foundations of our model in Section~\ref{sec:concepts}.
Our derivation is presented in Sections~\ref{sec:decay} and~\ref{sec:rate}, and
in Section~\ref{sec:implications} we discuss broader implications of our approach.

\section{Model Foundations}\label{sec:concepts}
Addressing virtual particles is usually reserved for field theories and second quantization.
To apply the tools of standard quantum mechanics to bound systems, we will consider
the following conceptual foundations and assumptions.

\subsection{Binding Interactions}\label{sec:bindInt}
We assume that the interactions among the components of bound systems and with the background fields
are no different from scattering interactions or interactions with measuring devices.\footnote{
    As in \cite{haagSharpnessLocalizationIndividual2013},
    we consider local observables in Quantum Field Theory to
    be analogues of the detectors of standard Quantum Mechanics.
}
If components are not already in the space of preparable bound states,
upon interacting they will be projected onto such states.
Whether measurement or scattering interactions are instantaneous or evolve by dynamical rules is
not presently known.
We note that
assuming finite measurement times, particularly times sensitive to interaction coupling 
type~\cite{brasilHowMuchTime2013},
may reveal subtleties in decay processes.
For simplicity, however, in the present work we will assume instantaneous projections.

We will model sequences of $N$ interactions,
and we will represent the state of a bound system immediately following the $n$th interaction
from a sequence with the density operator $\rho^n$.
All or at least a significant subset of the $N$ interactions will be binding interactions
involving the components of the bound system.
Because a bound system is presumably bound at all times, we will assume the continuous limit of
a sequence of discrete interactions.

The continuous limit of instantaneous projections that we adopt is the same limit assumed for the
QZE~\cite{misraZenosParadoxQuantum1977}.
Distinguishing between survival-as-prepared and non-decay, as described next, is an 
essential step toward understanding
why the QZE does not apply to systems with uncontrolled internal dynamics.\footnote{
    The explanation will be completed in the text following equation~\eqref{eq:quickres}.
}

\subsection{Survival-as-Prepared vs. Non-Decay}\label{sec:nondecay}

The modern view of a composite, bound system having internal dynamics mediated by
a teeming bundle of virtual particles 
suggests that an excited bound system will have a large number of possible internal configurations,
each described by a different metastable state in the space of undecayed states.
We must therefore distinguish the probability for \textit{survival-as-prepared}
from the probability for \textit{non-decay}.
In particular, the Hilbert space for such an excited state is not spanned by a single vector,
but is a product of a possibly changing number of spaces representing a possibly changing
number of internal quanta.
Furthermore, each space in the product represents an evolving sub-system and therefore
may be spanned by multiple discrete or continuous state vectors.

Let $n=0$ label the interaction when a bound system is initially prepared in an excited state.
For a system having two primary components, $a$ and $b$, and virtual quanta 
$\alpha$, $\beta$, etc, the density operator immediately after preparation is
\begin{equation}
    \rho^0 = \rho_a^0\otimes\rho_b^0\otimes\rho_{\alpha}^0\otimes\rho_{\beta}^0\otimes\ldots
\end{equation}
Immediately after the first binding interaction, or $n=1$ in the sequence, if the
system did not decay from the excited state and if the virtual quanta $\alpha$ and $\beta$ persisted,
it is represented by the metastable state operator
\begin{equation}
    \rho^1 = \rho_a^1\otimes\rho_b^1\otimes\rho_{\alpha}^1\otimes\rho_{\beta}^1\otimes\ldots
\end{equation}
If $\rho^{1}=\rho^0$,
then the system will have both survived as prepared and not decayed.
If, however, $\rho^{1}\neq\rho^0$, then the system will not have decayed,
but it also will not have survived as initially prepared.

\subsection{Disordered Internal Environment}\label{sec:disorder}
To model the presumed disorder of the uncontrolled, internal environment of a bound system,
we will use a stochastic sequence of interactions.

As an illustration, consider an analogous sequence in a controlled experiment, where interactions
are active measurements rather than scattering events:
an electron's spin is measured along a sequence of axes, where at each measurement
in the sequence the axis is chosen to point in a random direction.
In such a sequence of measurements, the probability is nearly zero that consecutive measurements will
leave the electron in states that commute.
Similar sequences with non-commuting observables have been studied~\cite{wangOperatorQuantumZeno2013,liQuantumZenoEffect2013},
but by focusing on decay rather than system dynamics, we will find a different result.

In the context of a bound system and using the notation of the previous subsection,
a component of the system labeled $a$ will be represented throughout the sequence
of binding interactions by $\rho_a^n$.
As a consequence of the uncontrolled and disordered interior of an unstable particle,
we will assume $[\rho_a^n,\rho_a^{n+1}]\neq 0$ for almost all interactions.
This is natural to imagine for a component interacting, say,
with a sequence of background fields,
each carrying an angular momentum randomly aligned in space.

\section{Exponential Decay}\label{sec:decay}
Let the operator $\Lambda_u^i$ project onto a possible metastable state labeled $i$, of the 
various components of a preparable, bound system.
By the assumption in Section~\ref{sec:bindInt},
if the system is found to be in an undecayed state upon the $n$th interaction,
corresponding to the observable $\Lambda_u^n$, we will represent it by a density operator $\rho^n$
from a space of undecayed states.
By the assumption in Section~\ref{sec:nondecay},
undecayed states in the sequence may differ from each other internally.
If the system is found undecayed upon the $n$th interaction in the sequence,
the probability that it will remain undecayed at the $(n+1)$th interaction, occurring 
a duration in time $\delta$ later, is
$\Tr\left(\Lambda_u^{n+1}\rho^n(\delta)\right)$.

Every member of an experimental ensemble will be prepared in a unique
sequence of metastable states before finally decaying.
Because we are concerned only with decay and not with the internal states themselves,
we will drop the index $n$ from the operators.
For our purposes, then,
$\Lambda_u$ represents the coarser observable that the bound system is undecayed.
The index $n$ should also label the durations between interactions, $\delta$,
but in the continuous limit, they will all go to zero, and we
have omitted the $n$ for clarity.

Once prepared initially,
the probability for the system to continue to be found
in the excited state throughout the interaction sequence is
$\Tr\left(\Lambda_u\rho(\delta)\right)^N$.
As noted by the authors of~\cite{misraZenosParadoxQuantum1977}, this product
must be distinguished from the typical probability for a single
measurement interaction.
In quantum kinematics it is known as a compound
measurement~\cite{schwingerQuantumKinematicsDynamics2018},
and here it is the probability,
given a sequence of $N$ equally spaced interactions over a total time $N\delta$, 
that each interaction finds the system undecayed after the
preceding interaction found it undecayed as well.

The continuous internal interactions enabling continuous binding
correspond to the limit $N\rightarrow\infty$ while holding $N\delta\equiv t$ finite.
The probability as a function of time, for internal interactions to find the bound system undecayed, is
\begin{equation}\label{eq:firstp}
    \mathcal{P}_u(t=N\delta) = \lim_{N\rightarrow\infty} \Tr\left(\Lambda_u\rho(\delta)\right)^N,
\end{equation}
after it was prepared in an excited state for the first time at $N\delta=0$.
Expanding the individual probabilities in the sequence for small $\delta$ gives
\begin{equation}\label{eq:pexpant}
\Tr\left(\Lambda_u\rho(\delta)\right)=
1 - \delta/\tau + \mathcal{O}\left(\delta^2\right),
\end{equation}
where $\tau^{-1}$ is given by
\begin{equation}\label{eq:decrate}
\tau^{-1} = - \frac{d}{d\delta} \Tr\left(\Lambda_u\rho(\delta)\right)
\bigg|_{\delta=0}.
\end{equation}
Using the time evolution 
$\rho(t)=e^{-iH(t)/\hbar}\rho(0)e^{iH(t)/\hbar}$,
where $H$ is the self-adjoint Hamiltonian operator of the system,
equation~\eqref{eq:decrate} becomes
\begin{equation}\label{eq:quickres}
\tau^{-1}=\frac{i}{\hbar}\Big( \Tr\big(\rho(\delta)\Lambda_u H \big)-
\Tr\big(\Lambda_u\rho(\delta) H \big) \Big)\bigg|_{\delta=0}.
\end{equation}

For the QZE,
in which a system is repeatedly projected onto the same, initially prepared state
in a controlled fashion,
$[\rho(\delta=0),\,\Lambda_u]=0$ by design,
and $\tau^{-1}=0$ for all measurements in the continuous sequence.
As a result, the
probability in equation~\eqref{eq:firstp} to find the system as prepared is
unity for all times.\footnote{
    Commutation is often assumed also for the Quantum Anti-Zeno Effect
    whereby less frequent projections drive a system from its prepared state faster
    than exponentially~\cite{kofmanAccelerationQuantumDecay2000}.
    The cited work also argues against the QZE in radioactive decay, noting that the frequency
    required for QZE freezing would be so high that,
    by the energy-time uncertainty relation,
    interactions could couple to high-energy channels for particle creation and destroy
    the excited system.
    Interestingly, while our approach and conclusion differ,
    passive virtual particle creation is
    of course integral to our mechanism for continuous interaction.
}

To model the uncontrolled internal interactions of an excited bound system, however,
we appeal to the assumption of a disordered internal environment from Section~\ref{sec:disorder},
with the consequence that $[\rho(\delta=0),\,\Lambda_u]\neq 0$ throughout the sequence.
With a nonzero $\tau^{-1}$ in equation~\eqref{eq:pexpant},
our model predicts exponential decay:
\begin{equation}\label{eq:limcalc}
    \mathcal{P}_u(t) = \lim_{N\rightarrow\infty} \Tr\left(\Lambda_u\rho(\delta)\right)^N =
e^{-t/\tau}.
\end{equation}
Note that exponential decay follows from the kinematics of
continuous internal interactions and is therefore general and independent of a bound system's dynamics.

This shift from the QZE is not a surprise conceptually.
It is known that varying projections can
force a system to change~\cite{schulmanContinuousPulsedObservations1998}.
Simply put: an electron prepared in a state spin-up along the $z$ axis will clearly not
survive as prepared if its spin is measured next along the $x$ axis.

\section{Decay Rate}\label{sec:rate}
A straightforward calculation of the decay rate in~\eqref{eq:decrate}
results in Fermi's Golden Rule (FGR).
A particular strength of the present derivation is that it illuminates the assumptions
required to apply various forms of the FGR.
To connect with this well-known formula and to compare with standard
textbook derivations, we will shortly change to bra-ket notation
for the derivation and treat a non-relativistic system.
This is of course not required.

As discussed above, $\Lambda_u$ may project onto different metastable
states for each interaction in the sequence.
Given this and the fact that the internal workings of unstable bound systems
are presently not well understood, writing the $\Lambda_u$ will be difficult.
Fortunately we can work instead with $\Lambda_d$, which we define to project onto
decayed states of the system.
Then $\Tr\left(\Lambda_d\rho(\delta)\right)$ is the probability to find the system
decayed, and it is related to the undecayed probability by
\begin{equation}\label{eq:probrel}
\Tr\left(\Lambda_d\rho(\delta)\right)= 1 - \Tr\left(\Lambda_u\rho(\delta)\right).
\end{equation}
With
$\frac{d}{dt}\mathcal{P}_u\big(\rho(t)\big)=
-\frac{d}{dt}\mathcal{P}_d\big(\rho(t)\big)$ we can rewrite the decay rate~\eqref{eq:decrate}
as a function of the decay probability:
\begin{equation}\label{eq:decraten}
\tau^{-1} = \frac{d}{d\delta} \Tr\left(\Lambda_d\rho(\delta)\right)
\bigg|_{\delta=0}.
\end{equation}

The shift in focus from $\Lambda_u$ to $\Lambda_d$ and the decayed states
has potentially significant implications for quantum engineering, which will be discussed
in the next section.
For now, making the shift allows us to consider
the asymptotically free decay products and use the standard
machinery of scattering theory and a perturbation approach to proceed.

The operator $\Lambda_d$ is
\begin{equation}
\Lambda_{d} = \sum_\eta\int\,d^2\Omega \, dE\,
\lambda(E,\Omega,\eta)\,\Lambda_{E,\Omega,\eta},
\end{equation}
where
$\Lambda_{E,\Omega,\eta}=
\ketbra{\psi_{E,\Omega,\eta}}{\psi_{E,\Omega,\eta}}$
are projection operators onto the infinitesimal subspaces representing the
states of the decayed system,
and $\lambda(E,\Omega,\eta)$ is the density of these states.
$E$ is the energy, and $\eta$ represents the excitation level of the
system, as well as any other quantum numbers required to specify its state.
$\Omega$ describes the direction of momentum of the decay products.
To connect with the simplest form of FGR,
we assume spherical symmetry and an isotropic density of final states.
After integrating over $\Omega$, we are left with
\begin{equation}\label{eq:obsexp}
\Lambda_{d} = \sum_\eta\int\,dE\, \lambda(E,\eta)\,\Lambda_{E,\eta},
\end{equation}
where 
\begin{equation}
    \Lambda_{E,\eta}=\ketbra{\psi_{E,\eta}}{\psi_{E,\eta}}.
\end{equation}

We partition the total Hamiltonian as
$H=H_0+V$, where $V$ represents interactions,
and $H_0$ is the free Hamiltonian for all components of the system after decay.
$\Lambda_d$ projects onto the lower energy states of the bound system
as well as any emitted decay products,
so $\ket{\psi_{E,\eta}}$ are dimensionless eigenkets of $H_0$.
Energy eigenkets of $H_0$ are $\ket{E,\eta}$, and
eigenkets of $H$ are the Lippmann-Schwinger kets, $\ket{E,\eta^+}$, 
where~\cite{lippmannVariationalPrinciplesScattering1950}
\begin{equation}\label{eq:L-S_kets}
\ket{E,\eta^+}=\ket{E,\eta}+
\lim_{\epsilon\rightarrow 0}\frac{1}{(E-H_0) + i\epsilon}V
\ket{E,\eta^+}.
\end{equation}
Here we have dropped another $\Omega$ because we will work in the rest
frame of the excited system.

Using equation~\eqref{eq:obsexp}
and writing the density operator of the bound system at $\delta=0$ as
$\rho(0)=\ketbra{\phi}{\phi}$,
the decay rate from~\eqref{eq:decraten} becomes
\begin{equation}\label{eq:draterwrte}
\tau^{-1} =
\frac{2}{\hbar}\textnormal{Im}
\sum_{\eta_f} \int \, dE_f\,
\bra{\psi_{E_f,\eta_f}}V\ket{\phi} \nonumber 
\braket{\phi}{\psi_{E_f,\eta_f}}\,\lambda(E_f,\eta_f),
\end{equation}
where we have used $[\Lambda_d,H_0]=0$.
We have added the subscript $f$ to distinguish the energy and quantum numbers of the final state.
$\bra{\phi}$ is the bound state vector, so we expand it in terms of the total Hamiltonian
eigenkets, $\ket{E,\eta^+}$.
Using~\eqref{eq:L-S_kets}
once, we have for the decay rate,
\begin{align}\label{eq:intermediate1}
\tau^{-1}  = & \lim_{\epsilon\rightarrow 0}
\frac{2\,\textnormal{Im}}{\hbar}
\sum_{\eta,\eta_f} \int \, dE\, dE_f\, 
\lambda(E_f,\eta_f) \nonumber \\
 &\bigg( \bra{\psi_{E_f,\eta_f}}V\ket{\phi}
\braket{\phi}{E,\eta^+}\braket{E,\eta}
{\psi_{E_f,\eta_f}}\nonumber \\
 &+ \bra{\psi_{E_f,\eta_f}}V\ket{\phi}
\braket{\phi}{E,\eta^+} 
\bra{^+E,\eta}
V\frac{1}{E-H_0-i\epsilon}\ket{\psi_{E_f,\eta_f}} \bigg).
\end{align}

The first term on the right hand side of~\eqref{eq:intermediate1} contains
$\braket{E,\eta}{\psi_{E_f,\eta_f}}$,
which is proportional to the
Kronecker delta $\delta_{\eta\eta_f}$, where $\eta_f$ represents, among
other things, the final excitation level of the system after decay.
Because $\braket{\phi}{E,\eta^+}$ is the wave function of
the excited system, the first term vanishes.
We are then left with
\begin{equation}\label{eq:form1}
\tau^{-1}=\frac{2\pi}{\hbar}
\sum_{\eta,\eta_f}\int dE\, dE_f\,
\bra{\psi_{E_f,\eta_f}}V\ket{\phi}
\braket{\phi}{E,\eta^+} 
\bra{^+E,\eta} V\ket{\psi_{E_f,\eta_f}}\,
\lambda(E_f,\eta_f) \delta(E-E_f).
\end{equation}

While exponential decay results in our model from binding interactions,
it is clear from~\eqref{eq:form1} that contributions to the decay rate
come from interactions coupling the excited systems to the decay products,
as expected.
The delta function in~\eqref{eq:form1} enforces energy conservation.
The final energy, $E_f$, includes the energy of all decay products.
One can write, for instance, $E_f = E_{\eta,f}+\hbar\omega_f$, where
$\omega_f$ is the angular frequency of an outgoing photon.
Equation~\eqref{eq:form1}
is our most general result for the decay rate, $\tau^{-1}$,
of an excited system.
Recall that, for brevity, we did assume an isotropic density of final states
above for equation~\eqref{eq:obsexp}, and for a non-isotropic
density equation~\eqref{eq:form1} would change in straightforward way.

Until now we have avoided specifying the state of the excited system at
preparation, but when the excited system is prepared with sharp values,
$E_i$ and $\eta_i$, such that 
\begin{equation}\label{eq:fapprx}
\sum_\eta\int dE\, dE_f \ket{E,\eta^+}
\braket{^+E,\eta}{\phi}\delta(E-E_f)
= \int dE_f \ket{\phi_{E_i,\eta_i}}\delta(E_i-E_f),
\end{equation}
we can write~\eqref{eq:form1} in infinitesimal form:
\begin{equation}\label{eq:form2}
    \tau^{-1} = \frac{2\pi}{\hbar}\sum_{\eta_f}\int dE_f 
               |\bra{\psi_{E_f,\eta_f}}V\ket{\phi_{E_i,\eta_i}}|^2
\lambda(E_f,\eta_f)\delta(E_i-E_f).
\end{equation}

For a channel labeled by the initial and final
internal quantum numbers $\eta_i$ and $\eta_f$,
we have for the decay rate
\begin{equation}\label{eq:partdecrat}
\tau_{\eta_i\rightarrow\eta_f}^{-1} = \frac{2\pi}{\hbar}
|\bra{\psi_{E,\eta_f}}V\ket{\phi_{E,\eta_i}}|^2 \lambda(E,\eta_f),
\end{equation}
where the notation $E_i=E_f\equiv E$ reflects energy conservation.
Thus we have derived Fermi's Golden Rule for a time-independent decay rate.
Though it is obscured in the notation of~\eqref{eq:fapprx},
$\ket{\phi_{E,\eta_i}}$ is proportional to a Lippmann-Schwinger ket,
which can be iteratively expanded to obtain the Born series.

In the definition of $\Lambda_d$, we sum over $\eta_f$ and implicitly over
different decay channels.
The derivation does extend naturally to distinct decay channels, however.
Partition the interaction potential, $V$, into two parts:
\begin{equation}\label{eq:vparts}
V=V_k + V_\ell.
\end{equation}
If there is a final state, $\ket{\psi_{E_{fk},\eta_{fk}}}$,
such that
\begin{equation}\label{eq:sep1}
\int dE \braket{\phi}{E,\eta^+} 
\bra{^+E,\eta} V_k\ket{\psi_{E_{fk},\eta_{fk}}}
\delta(E-E_{fk}) \neq 0,
\end{equation}
and
\begin{equation}\label{eq:sep2}
\int dE \braket{\phi}{E,\eta^+} 
\bra{^+E,\eta} V_\ell\ket{\psi_{E_{fk},\eta_{fk}}}
\delta(E-E_{fk}) = 0,
\end{equation}
then the decay rate for a channel labeled $k$ is
\begin{equation}\label{eq:partdecrata}
\tau_k^{-1} = \frac{2\pi}{\hbar}
|\bra{\psi_{E,\eta_{fk}}}V_k\ket{\phi_{E,\eta_i}}|^2
\lambda(E,\eta_{fk}),
\end{equation}
and similar for $\ell$.
This generalizes to any number of partitions into channels, and
it is straightforward to show that the total decay rate
is the sum of the rates for various channels, as expected.
When there are two available decay channels, as indicated
in~\eqref{eq:vparts}, the non-decay
probability as a result of continuous internal interactions is
\begin{equation}
\mathcal{P}_u(t) =
e^{-t/\tau_k}\, e^{-t/\tau_\ell},
\end{equation}
as verified by experiments.

\section{Discussion}\label{sec:implications}
Perhaps our result most significant for the development of quantum technology and engineering
is the distinction between the interactions driving exponential decay and those
interactions contributing to the decay rate.
This insight followed the shift in focus from 
non-decay with $\Lambda_u$ in equation~\eqref{eq:decrate}
to decay with $\Lambda_d$ in equation~\eqref{eq:decraten}.
Specifically, the potential $V$ coupling the states of the excited systems
to the states of the decay products appears in FGR
and controls the decay rate, but it does not represent the internal binding interactions
that lead to the exponential character of decay.
One consequence is that engineering or even
eliminating decay channels may prolong overall lifetimes, but excited
bound systems will continue to suffer disordered internal interactions and evolve incoherently.

An interesting application of our model will be to investigate
the apparent continuum of observed decay phenomena mentioned in the Introduction.
Our development clarifies the
difference between the QZE for controlled systems that survive as prepared
and the disordered internal environment of bound systems leading to exponential decay.
These two types of system represent opposite ends of a continuum of decay phenomena.
Furthermore, observations of late-time deviations from exponential decay for dissolved organic molecules,
that depend on the environment (solvent) and molecule length hint at a
middle ground~\cite{rotheViolationExponentialDecayLaw2006}.
Recent results~\cite{scheideggerCanIncreasingSize2025}
indicate that longer organic molecules better resist decoherence due to 
internal dynamics, consistent with our hypothesis that the interactions between system components
drive the decay in time.
Such organic molecules may be candidates to study the trade-off between internal interactions
and environmental coupling, and how decay may or may not be affected or controlled.

Similarly, for uncontrolled, metastable bound systems, such as excited nuclei or atoms,
we modeled continuous binding as a limit of instantaneous projections.
For such systems, observed deviations from exponential decay (or the lack thereof)
could imply (or constrain) finite timescales for scattering interactions.

Furthermore, while we assume projective measurements,
quantum systems undergoing continuous, non-demolition probes of non-commuting
observables have been observed~\cite{hacohen-gourgyQuantumDynamicsSimultaneously2016},
and sequences of
unitary interactions with varying ancillary systems have been studied~\cite{grimmerOpenDynamicsRapid2016}.
These would provide interesting avenues to extend our model to non-projective internal interactions,
especially in conjunction with controlled environmental interactions.

Of more fundamental interest is a resolution to the arguments
regarding the stability of matter and the impossibility of
a decaying system having the Lorentzian energy wave function corresponding to 
exponential decay~\cite{l.a.khalfinContributionDecayTheory1958}.
With our model, this limitation does not apply, because a bound, decaying
system is not represented by
a continuously evolving state vector throughout the decay process.
In the absence of interactions,
$\phi(E)$ need not be a Lorentzian, and the system can be bounded from below in energy.
Rather, the Lorentzian line shape
emerges as a consequence of continuous interactions.

\section{Conclusion}
We assume the internal binding interactions of bound systems are no different from 
scattering interactions or measurement interactions that project systems onto
preparable states.
To model continuous binding, we assume the continuous limit of instantaneous projections,
as done originally also for the QZE.
Because bound systems have changing internal configurations, however,
we derive an exponential non-decay probability at all times.
The result follows from kinematics and therefore applies to any bound system with internal
dynamics and multiple internal, non-commuting configurations.
If continuous internal interactions over a finite $t=N\delta$
are realized in nature, then they are a candidate to explain 
quantum mechanically the normally empirically or
approximately~\cite{weisskopfBerechnungNaturlichenLinienbreite1930} derived exponential decay law.

Furthermore, calculation of the decay rates results in a
straightforward derivation of Fermi's Golden Rule for any number of decay channels.
While we believe our derivation is particularly transparent and informative,
there have been operationally similar derivations based also on a time-dependent perturbation
approach (for example,~\cite{weissbluthAtomsMolecules2012}).
Such derivations are complicated by justifying a time-independent decay rate, however.
As seen in equation~\eqref{eq:decrate}, here time independence is a consequence of continuous interactions.

As far as we know, this is the first approach that results in predictions of
exponential decay together with a revealing derivation of
Fermi's Golden Rule for time-independent decay rates.
Our framework provides insight into the types of interactions that can cause decay,
and suggests an approach to study the continuum of decay phenomena,
with systems controlled via the QZE on
one end and uncontrolled atoms and nuclei on the other.
We expect these insights will be significant for the development of quantum technology.

Finally, the framework avoids various foundational difficulties with exponential decay.
The objection from the boundedness of Lorentzian functions does not apply because the
lineshape is emergent and not representative of the energy wave functions of states.
Similarly, the exponential time evolution results from the kinematics of constant interactions 
rather than the time evolution properties of a state vector.

\end{document}